\relax
\documentclass[letterpaper]{article}
\usepackage{aaai17}
\usepackage{times}
\usepackage{helvet}
\usepackage{courier}
\usepackage{url}  
\usepackage{graphicx}
\usepackage{subfig}
\usepackage{array,multirow}
\usepackage{footnote}
\begin{document}

\title{Fertility and its Meaning: Evidence from Search Behavior\thanks{This is a preprint of a short paper accepted at ICWSM'17. Please cite that version instead.}}
\author{Jussi Ojala\\Aalto University\\jussi.k.ojala@aalto.fi \And Emilio Zagheni\\University of Washington\\emilioz@uw.edu \And Francesco C. Billari\\Bocconi University\\billari@unibocconi.it\And Ingmar Weber\\Qatar Computing Res.\ Inst., HBKU\\iweber@hbku.edu.qa}

\maketitle

\begin{abstract}
Fertility choices are linked to the different preferences and constraints of individuals and couples, and vary importantly by socio-economic status, as well by cultural and institutional context. The meaning of childbearing and child-rearing, therefore, differs between individuals and across groups. In this paper, we combine data from Google Correlate and Google Trends for the U.S.\ with ground truth data from the American Community Survey to derive new insights into fertility and its meaning. First, we show that Google Correlate can be used to illustrate socio-economic differences on the circumstances around pregnancy and birth: e.g., searches for ``flying while pregnant'' are linked to high income fertility, and ``paternity test'' are linked to non-marital fertility. Second, we combine several search queries to build predictive models of regional variation in fertility, explaining about 75\% of the variance. Third, we explore if aggregated web search data can also be used to model fertility trends. 
\end{abstract}

\section{Introduction}

Having a child has far-reaching consequences in the life of individuals. Fertility and its trends shape populations and therefore societies. A vast literature discusses fertility see e.g, \cite{Balbo2013} and its meaning for individuals, couples and societies. Given the far-reaching consequences of parenthood, gathering information is crucial for individuals who decide to have (or to prevent having) a child, the circumstances in which a to give birth, and how to rear a child. Planning has become more central and in the U.S.\ unplanned pregnancies have been recently declining \cite{Finer2016}. Information available online is likely to contribute to the process of information gathering, and therefore, to fertility decisions, and it might also be important in shaping the decline in unplanned pregnancies in the U.S. \cite{Kearney2015}. More generally, information gathering online is likely to occur even in cases of unplanned pregnancies and birth, and after birth, for child-rearing.

Google searches are the most obvious instances of information gathering online. As there are important socio-economic differences in fertility outcomes--for instance concerning the prevalence of unplanned pregnancies but also of ``parenting style''--we expect online information gathering on fertility and child-rearing to differ between groups and between contexts.

We explore the use of Google Correlate and Google Trends data, linked to ground truth estimates from the American Community Survey, to (i) detect evidence for different contexts surrounding different ``meanings'' of fertility; (ii) model regional variation across states for different fertility levels; (iii) track temporal changes in fertility across time. We find that Google search provides strong signals for (i) and (ii) but not for (iii).

\section{Related Work}
Though Google Trends has been publicly available since May 2006, its use for tracking real-life quantities became popularized through the creation of Google Flu \cite{detectingflu}, a service for ``now-casting'' the prevalence of flu based on search volumes for particular queries. Other popular applications have been related to tracking economic indicators \cite{ECOR:ECOR809}. A review of papers, though mostly focusing on applications in health can be found in \cite{SurveyPaper}. Limitations due to Google's ``black box'' and due to changes in user behavior over time are discussed in \cite{gftfail}.


The most closely related work is \cite{billariPAA2013} where the authors use temporal variation in search volumes to forecast fertility rates for U.S.\ states. Our approach is different as we focus on \emph{spatial} variation, on the meaning and context of births, and make use of Google Correlate. Another relevant study using Google Correlate is \cite{CorrelationStats} which shows that the search terms Google Correlate lists for ``birth rates" and ``infant death rates" in different states are statistically meaningful.


Also related to fertility, Reis and Brownstein \cite{Reis2010} show that the volume of Internet searches for abortion is \emph{inversely} proportional to local abortion rates and directly proportional to local restrictions on abortion.  

\section{Data Collection}
The departing point for our data collection is U.S.\ Census Bureau fertility data\footnote{\url{https://factfinder.census.gov}}, part of the American Community Survey (ACS) \footnote{\url{https://goo.gl/qr0dyu}}. We obtain state-level estimates, including Washington D.C., for nine different fertility-related variables, listed in Table~\ref{tab:variables}. This data was obtained for the years from 2010 to 2015 and relates to women that have given birth during the corresponding year. The two income related groups are defined with respect to the ``poverty level''\footnote{\url{https://goo.gl/KFizlA}}.

\begin{table}[ht]
\begin{tabular}{lcccc}
Fertility rates & Abbrv. & Av. & Med. & Std. \\ \hline
General, age 15-50 & Gen. & 54  & 54 & 6.5\\
Marital: unmarried & MSnot  &19  & 19& 3.3 \\
Marital: married & MSyes  & 35  & 34 & 6.5\\
Young, age 15-19 & Teen & 2.6 & 2.5 & 0.8 \\
Old, age 35-50 & Old & 10 & 10 & 2.0\\
Education: univ. & Ehigh &17& 16 & 3.6 \\
Education: other & Elow &37 & 37 & 6.6 \\
Poverty: up to 100 & Poor & 14 & 14 & 3.2 \\
Poverty: from 200 & Rich &28 & 27& 5.0 \\
\end{tabular}\caption{The list of the nine fertility-related variables used along with descriptive statistics for 2015 on between-states variation. The values are multiplied by 1,000, and normalized by dividing the counts of certain types of births by the total number of women aged 15-50 in each state.
}\label{tab:variables}
\end{table}

Google Trends data are normalized with respect to Google search volume. We therefore normalized ACS data by dividing the number of births in the reference group by the total number of women aged 15-50.\footnote{For convenience, we used the number of women aged 15-50 rather than the state's total population as (i) the two are strongly correlated, and (ii) the latter was not part of the aggregate fertility related data exported by \url{http://factfinder.census.gov}.} These nine quantities, which one could view as ``fertility intensities'', no longer represent intuitive percentages but they are expected to correlate more naturally with the normalized Google search volume.

We then chose the year 2015 as our reference year and uploaded the nine series of 2015 fertility intensities across the 51 states to Google Correlate\footnote{https://www.google.com/trends/correlate/}. For each of the series, Google provided a list of the top 50 search terms most strongly correlated in terms of their spatial search intensity distribution.  As an example, see \url{https://goo.gl/a8Sf65} for the results for the General Fertility, with the .csv file available at \url{https://goo.gl/2EXnie}. The search intensity measures the total percentage of Google search volume in a U.S.\ state that is made up by the search term. Along with the list of terms we also obtained (i) the Pearson r correlation and (ii) a z-score (mean 0, std.\ dev 1.0) normalized series of the 51 search intensities.

We  manually post-filtered the 
correlated search terms to create lists of at most five search terms. The goal of this step was to remove spurious correlations, where unrelated search terms happen to be correlated by chance. Instead, we created a list of terms that could at least have some plausible link to fertility, sexual activity (including sexually transmitted infections), caring for a baby, or family formation. 

The selection proceeded by going down the list of correlated search terms in decreasing order of correlation. In case the candidate search group included both plural and singular for one search term, the less correlated search term was removed from the selected set. The selection stopped when either five search terms had been identified or when the end of the list was reached. This procedure resulted in a total of 28 search terms and two of the nine fertility related variables had to be dropped as none of the correlated search terms passed our filtering. Concretely, fertility rates for `poor' and `old' were removed. These two categories mostly had correlated search terms related to (i) real estate and cars, and (ii) U.S.\ immigration and international travel, respectively. Table~\ref{tab:selected_terms} shows the final list of 28 search terms together with their Pearson $r$ correlation with the 2015 values.

\begin{table*}[ht]
\centering
\setlength{\tabcolsep}{2pt}
\scriptsize 
\begin{tabular}{ccccccc}
Gen & MSnot & MSyes & Teen & Elow & Ehigh & Rich \\ \hline
pregnancy workout (.88)  & chlamydia and gonorrhea (.81) & nursing cover (.84)  & names from the bible (.84)  & how to potty (0.90)  & crib reviews (.81) & post pregnancy (.75)   \\
baby tummy (.87)  & biblical names (.77) & jogging stroller reviews (.83)  & baby in the womb (.81)  & how to potty train (0.90)  & week 37 (.74) & chicco key (.73)   \\
baby constipation (.86)  & treatment for chlamydia (.77) & double jogging stroller (.82)  & None  & potty train (0.88)  & None &  baby stuffy nose (.73) \\
increase breast milk (.85)  & paternity test (.76) & nursing pads (.82)  & None  & uddercovers.com (0.86)  & None & flying while pregnant (.72)    \\
baby trend (.85)  & transmitted disease (.76) & jogging stroller (.81)  & None  &  None & None & baby card (.71)    \\
\end{tabular}\caption{(Up to) Top five terms selected from the top 50 correlated terms on Google Correlate for the seven different fertility measures with at least one baby/birth-related search term. The correlation with the 2015 fertility rates is shown in parentheses.}\label{tab:selected_terms}
\end{table*}

%
\section{Results}

%
\subsection{Evidence for Different Meanings of Fertility}
Table~\ref{tab:selected_terms} provides some evidence that search term reflect the different meanings of birth for different socio-economic groups. For example, the `Rich' birth rates correlate with queries related to air travel or high-end child car seats\footnote{Chicco Key is a child car seat system retailing at roughly twice the price of budget variants.}. 
The unmarried birth rate on the other hand correlates with search terms related to sexually transmitted infections and paternity tests. Both teen and unmarried birth rates correlated with references to biblical names.
This evidence fits with previous analyses that have documented that religiosity is positively correlated with teen birth rates in the U.S.~\cite{Strayhorn2009}, as well with research showing that state-level measures of religiosity are positively correlated with Google searches for sexual content online \cite{MacInnis2015}.


%
\subsection{Modeling Spatial Variation}
The previous section demonstrates that Google Correlate can indeed pick up search terms related to different meanings attached to fertility. We now test whether combinations of these search terms can be used to model regional variation in these rates across the 51 states.

We use a standard regression setup where for a given state $i$ the dependent variable $y_i$ is a particular fertility rate as measured in 2015. See Table~\ref{tab:variables} for the full list of nine fertility rates of which seven corresponded to at least one meaningful search term. The feature matrix $X$ contains one row for each state and one column for each of the (up to) five search terms in Table~\ref{tab:selected_terms}. These columns come from Google Correlate\footnote{See \url{https://goo.gl/itVVTj} for an example for the general fertility rates.} and have been pre-normalized to z-scores across the 51 states. We add one additional column to allow the fitted linear models to contain an offset.

For each of the seven pairs of ($y$, $X$) we then evaluate the predictive performance of (i) a model that uses the state-wide average as a constant prediction; (ii) a single-term linear regression; (iii) a linear regression with all search terms; (iv) a Lasso generalized linear model with regularization \cite{tibshirani1996regression}. As performance measures we report (i) the Pearson $r$ correlation; (ii) the root-mean-square error (RMSE); (iii) the symmetric mean absolute percentage error (SMAPE). Performance was evaluated in a leave-one-out cross-validation setting. Table~\ref{tab:spatial_results} summarizes the results. 

\begin{table}[ht]
\centering
\scriptsize 
\begin{tabular}{ccccccccc}
Model & Metric & \rotatebox[origin=c]{90}{Gen} & \rotatebox[origin=c]{90}{MSnot} & \rotatebox[origin=c]{90}{MSyes} & \rotatebox[origin=c]{90}{Teen} & \rotatebox[origin=c]{90}{Elow} & \rotatebox[origin=c]{90}{Ehigh} & \rotatebox[origin=c]{90}{rich} \\ \hline
\multirow{3}{*}{\rotatebox[origin=c]{90}{Const.}} & $r$ & 0 & 0 & 0 & 0 & 0 & 0 & 0 \\
 & RMSE & 6.4 & 3.2 & 6.5 & .78 & 6.6 & 3.5 & 5.0 \\
 & SMAPE & 4.4 & 7.0 & 6.8 & 12.6 & 7.4 & 9.0 & 6.8 \\ \hline 
 \multirow{3}{*}{\rotatebox[origin=c]{90}{single}} & $r$ & .87 & .78 & .82 & .83 & .90 & .80 & .72 \\
 & RMSE & 3.2 & 2.0 & 3.7 & .44 & 2.9 & 2.1 & 3.5 \\
 & SMAPE &2.4 & 4.4 & 4.3 & 7.0 & 3.1 & 5.0 & 4.9 \\ \hline 
 \multirow{3}{*}{\rotatebox[origin=c]{90}{multi}} & $r$ & .89 & .77 & .83 & .85 & .93 & .86 & .79 \\	
 & RMSE & 2.9 & 2.1 & 3.6 & .42 & 2.4 & 1.8 & 3.1 \\
 & SMAPE & 2.1 & 4.3 & 4.4 & 6.4 & 2.7 & 4.3 & 4.1 \\ \hline 
 \multirow{3}{*}{\rotatebox[origin=c]{90}{Lasso}} & $r$ & .90 & .79 & .83 & .85 & .94 & .86 & .80 \\
 & RMSE & 2.8 & 2.0 & 3.6 & .42 & 2.3 & 1.8 & 3.0 \\
 & SMAPE & 2.0 & 4.2 & 4.4 & 6.4 & 2.6 & 4.3 & 4.0 \\ \hline
 \end{tabular}
\caption{Summary of the performance of the different regression models evaluated using leave-one-out cross-validation. SMAPE is in [\%], RMSE values are multiplied by 1,000. }\label{tab:spatial_results}
\end{table}

For each of the seven different fertility variables we then chose a \emph{single} model, i.e., either single-term, all-terms, or Lasso, based on predictive correlation value. Table~\ref{tab:spatial_results}  shows that, except for some ties, the LASSO models always performed best and was thus used in the next section for modeling temporal variation. 

The LASSO models did not use all variables. Concretely, it sparsified the models for `Gen' (removing `baby constipation'), for `MSnot' (removing `treatment for chlamydia' and `paternity test'), for `Elow' (removing `how to potty train') and for `rich' (removing `baby stuffy nose').

%
\subsection{Modeling Temporal Variation}
Above we followed a fairly standard ``train a model across space, and apply the model across space''. We now explore a novel ``train a model across space, and apply the model across time'' approach. Concretely, we use Google Trends data to track how the terms selected by the regularized LASSO model vary between 2015 and 2010 - going backwards in time - to see if the spatial model also ``predicts'', in retrospect, country-level changes in fertility rates across time. As there is data for only six years, training any type of vector auto regression model is doomed to fail. The hope is that by training across the 51 states we obtain a model that can then be successfully applied across time.

We do not attempt to model temporal changes at the state-level. The reason for this is that Google Trends, our data source for historic search volume, suffers from sparsity issues and, when restricted to a single state, fails to provide any data for the vast majority of the (state, search term) combinations that we would need to collect.

The spatial models were trained on data provided by Google Correlate which has been z-normalized, i.e., each search term had an average of 0 and a standard deviation of 1 across the 51 states. To obtain similar data across the years, we first summed the historic search volume obtained from Google Trends at the national level across the 12 months in each year. The series of six values, one for each year, was then z-normalized. This gave us data where each year, conceptually, is the same as a state.

\begin{figure}[ht]
\centering
\hspace*{-4mm} 
\includegraphics[height=4.0cm,width=.53\textwidth]{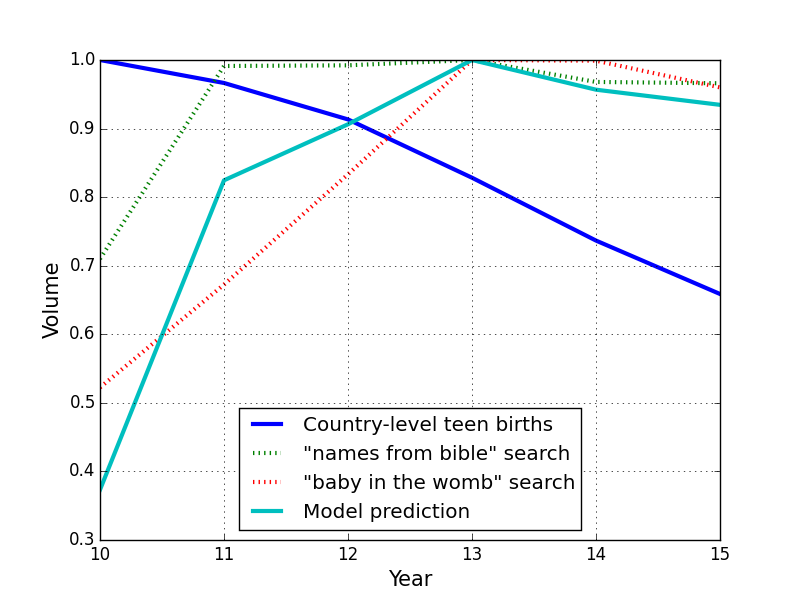}\caption{Temporal trend when applying the ``teen'' model across time. To emphasize the trend in volumes, values are rescaled to a maximum of 1.0. Surprisingly, there is a \emph{negative} correlation ($r=-0.65$) from 2015 back to 2010.}\label{fig:temporal_results}
\end{figure}

For each of the seven fertility variables, we then applied the corresponding spatial model to the 2015 national level data to obtain a reference prediction. The \emph{absolute} value of this prediction is not necessarily meaningful as the input values have been normalized in relation to previous years. However, we explored if the \emph{relative} trend of this prediction across the six years matched the ground truth. Table~\ref{tab:temporal} shows the results in terms of $r$ correlation across the six years. In most cases there was a \emph{negative correlation} due to the fact that whereas the birth rates have dropped over that period, the search intensity for baby-related topics has generally increased. Figure~\ref{fig:temporal_results} shows one such example.

\begin{table}[ht]
\scriptsize 
\centering
\begin{tabular}{cccccccc}
& Gen  & MSnot & MSyes & Teen & Elow & Ehigh & Rich \\ \hline
$r$ & -.574 & .019 & .257 &    -.657 &	-.436 &  -.266 & -.472 \\	
\end{tabular}\caption{Pearson $r$ correlation across 2010-2015 when using the spatial model to predict trends across time.}\label{tab:temporal}
\end{table}



\section{Discussion}

A key limitation when using Google Trends with fairly specific search terms such as ``increase milk supply'' is data sparsity. Even though Google Correlate indirectly provides relative state-level volume information, Google Trends only provides data for 22 of the 51 possible states. As explained in a previous section, this means we cannot obtain state-level historic volume information.

As Google does not provide a functionality to limit the analyzed data to a particular demographic group, it is hard to pick up trends within a particular sub-group. For example, the number of low-income births in a given year can be thought of as a product of one factor related to overall income and one factor related to birth rates. As we can only observe a single value, overall search volume, this makes it impossible to disentangle the two factors. We believe that this is the main reason that the terms correlated with general fertility rates look a lot more intuitive than the terms for low income fertility rates.

Whenever data is processed in an undisclosed manner this creates a black box. For example, we do not know which time period is used for Google Correlate  when looking for spatial correlations. In our analysis we use 2015 data as the reference year but other choices could be experimented with.

Finally, even though it did not give the expected results in our setting, our ``train across space, apply across time'' could be of interest for general nowcasting scenarios where the spatial resolution is higher than the temporal resolution. For demographic research this setting is not atypical as population statistics are often released only once per year, though at a fine-grained spatial resolution.

\section{Conclusions}
In this study we showed a relationship between fertility-related Google searches and socio-economic and demographic characteristics of births. The availability of information via the Web can affect demographic choices, and potentially reduce behavioral differences across groups. At the same time, Web searches partially reflect social structures.

Attitudes towards fertility, as well as practices related to child-rearing are difficult to measure with standard surveys, partially because of issues related to social desirability and conformity. Web searches are less likely to suffer from desirability biases, and offer a new perspective on the interests as well as lack of information that differentially characterize subgroups of the population. In this study, we provide a first sketch on the type of information that different subgroups are likely to seek in the context of childbearing.
We cannot infer individual-level features from aggregate searches, as we would run into the problem of ecological fallacy. Nonetheless, we showed that there is signal across socio-demographic groups and space in the U.S.

Our study has important implications for the field of demographic research, for population projections, and for improving the well-being of families and children. Finding queries that have solid relationships with fertility would help researchers understand the context and needs of particularly vulnerable groups, like teenage mothers. It would help sharpen public health information campaigns, and it would reveal the extent to which stigma is attached to various topics across space. In the context of developing countries that are still undergoing the fertility transition, web searches may provide relevant proxies about attitudes and offer information useful to predict the pace of future fertility reductions.
\bibliography{google_fertility}

\begin{thebibliography}{}

\bibitem[\protect\citeauthoryear{Balbo and others}{2013}]{Balbo2013}
Balbo, N., et~al.
\newblock 2013.
\newblock Fertility in advanced societies: A review of research.
\newblock {\em European Journal of Population / Revue europ{\'e}enne de
  D{\'e}mographie} 29(1):1--38.

\bibitem[\protect\citeauthoryear{Billari and others}{2013}]{billariPAA2013}
Billari, F.~C., et~al.
\newblock 2013.
\newblock Forecasting births using google.
\newblock http://paa2013.princeton.edu/papers/131393.

\bibitem[\protect\citeauthoryear{Choi and Varian}{2012}]{ECOR:ECOR809}
Choi, H., and Varian, H.
\newblock 2012.
\newblock Predicting the present with google trends.
\newblock {\em Economic Record} 88:2--9.

\bibitem[\protect\citeauthoryear{Finer and Zolna}{2016}]{Finer2016}
Finer, L.~B., and Zolna, M.~R.
\newblock 2016.
\newblock Declines in unintended pregnancy in the united states, 2008-2011.
\newblock {\em New England Journal of Medicine} 374(9):843--852.

\bibitem[\protect\citeauthoryear{Ginsberg and others}{2009}]{detectingflu}
Ginsberg, J., et~al.
\newblock 2009.
\newblock Detecting influenza epidemics using search engine query data.
\newblock {\em Nature} 457:1012--1014.

\bibitem[\protect\citeauthoryear{Kearney and Levine}{2015}]{Kearney2015}
Kearney, M.~S., and Levine, P.~B.
\newblock 2015.
\newblock Media influences on social outcomes: The impact of mtv's 16 and
  pregnant on teen childbearing.
\newblock {\em American Economic Review} 105(12):3597--3632.

\bibitem[\protect\citeauthoryear{Lazer and others}{2014}]{gftfail}
Lazer, D., et~al.
\newblock 2014.
\newblock The parable of google flu: Traps in big data analysis.
\newblock {\em Science} 343(14 March):1203--1205.

\bibitem[\protect\citeauthoryear{Letchford~A}{2016}]{CorrelationStats}
Letchford~A, Preis~T, M.~H.
\newblock 2016.
\newblock Quantifying the search behaviour of different demographics using
  google correlate.
\newblock {\em PLoS ONE 11}.

\bibitem[\protect\citeauthoryear{MacInnis and Hodson}{2015}]{MacInnis2015}
MacInnis, C.~C., and Hodson, G.
\newblock 2015.
\newblock Do american states with more religious or conservative populations
  search more for sexual content on google?
\newblock {\em Archives of Sexual Behavior} 44(1):137--147.

\bibitem[\protect\citeauthoryear{Nuti and others}{2014}]{SurveyPaper}
Nuti, S.~V., et~al.
\newblock 2014.
\newblock The use of google trends in health care research: A systematic
  review.
\newblock {\em PLOS ONE} 9(10):1--49.

\bibitem[\protect\citeauthoryear{Reis and Brownstein}{2010}]{Reis2010}
Reis, B.~Y., and Brownstein, J.~S.
\newblock 2010.
\newblock Measuring the impact of health policies using internet search
  patterns: the case of abortion.
\newblock {\em BMC Public Health} 10(1):514.

\bibitem[\protect\citeauthoryear{Strayhorn and Strayhorn}{2009}]{Strayhorn2009}
Strayhorn, J.~M., and Strayhorn, J.~C.
\newblock 2009.
\newblock Religiosity and teen birth rate in the united states.
\newblock {\em Reproductive Health} 6(1):14.

\bibitem[\protect\citeauthoryear{Tibshirani}{1996}]{tibshirani1996regression}
Tibshirani, R.
\newblock 1996.
\newblock Regression shrinkage and selection via the lasso.
\newblock {\em Journal of the Royal Statistical Society. Series B
  (Methodological)}  267--288.

\end{thebibliography}
\bibliographystyle{aaai}

\end{document}